\newcommand\bm[1]{\mbox{\boldmath$#1$}}

\def\sqr#1#2{{\vcenter{\vbox{\hrule height.#2pt
	\hbox{\vrule width.#2pt height#1pt \kern#1pt
	\vrule width.#2pt}
	\hrule height.#2pt}}}}

\documentclass[12pt,a4paper]{article}
\usepackage{color}

\begin{document}

\title{An approximate global solution of Einstein's equations for a differentially rotating compact body}
\author{A.\ Molina${}^{1,2}$ and E.\ Ruiz${}^3$\\[.5ex]
${}^1$\emph{Dep. de F\'\i sica Qu\`antica i Astrof\'\i sica 
}
\\
${}^2$\emph{Institut de Ci\`encies del Cosmos (ICCUB)}
\\
\emph{Universitat de Barcelona},
\\
\emph{Mart\'{\i} Franqu\`es 1, 08028 Barcelona, Spain}
\\
${}^3$\emph{Instituto  Universitario de F\'\i sica Fundamental y Matem\'aticas},
\\
\emph{Universidad de Salamanca},
\\
\emph{Plaza de la Merced s/n, 37008 Salamanca, Spain}.
}
\maketitle

\date{}

\begin{abstract}
We obtain an approximate global stationary and axisymmetric solution
of Einstein's equations which can be thought of as a simple star model: a self-gravitating perfect fluid ball with a differential rotation  
motion pattern. Using the post-Min\-kows\-kian formalism (weak-field approximation) and considering rotation as a perturbation 
(slow-rotation
approximation), we find approximate interior and exterior 
(asymptotically flat)
solutions to this problem in harmonic coordinates. Interior and exterior solutions are matched, in the sense 
described by Lichnerowicz, on the
surface of zero pressure, to obtain a global solution. The  resulting metric depends on four
arbitrary constants: mass density;
rotational velocity at $r=0$; a parameter that accounts for 
the change in rotational velocity through the star;  and  the star radius 
in the non-rotation limit. The mass, angular momentum,  quadrupole moment and other constants of the exterior metric are determined in terms of these four parameters.

\noindent
PACS number(s) 04.40.Nr, 04.20.Jb
\end{abstract}

\section{Introduction}
One of the regrettable facts about General Relativity is that, up to now, despite the numerous
exact solutions and modern methods to generate them, there has been no exact solution describing a rotating stellar model, i.e., a space-time corresponding
to an isolated self-gravitating rotating fluid in equilibrium, other than the
rotating disc of dust described by Neugebauer and Meinel in \cite{Neugebauer} and its generalization
for counter-rotating discs  \cite{Klein}. Although these infinitesimally thin disc solutions are useful
models for galaxies and accretion discs, they are far removed from a description of spheroidal sources,
which are the most common astrophysical objects.

Stellar models are built by matching an interior space-time describing the source and
the exterior space-time that encloses it. A candidate interior solution should correspond
to a stationary axisymmetric perfect fluid without extra symmetries and admit a zero
pressure surface. 

To our knowledge, for a long time the only candidates have been  the Wahlquist metric
\cite{Wahlquist}  and the generalization of this solution \cite{Senovilla}, both for a rigidly rotating perfect fluid with the equation of state: $\mu+3p=Ct.$  and the family of  differentially rotating solutions with the equation of state: $\mu=p+Ct.$   which have good properties \cite{Mars-Seno_2,Mars-Seno_3}; for instance, they verify the energy conditions, zero pressure surface, finite body and regular symmetry axis; but they have a small problem: there are  two singular points 
at the north and south poles. This could be avoided, and the work of Haggag \cite{Haggag} moved in this direction, but again, that solution has a singularity: a Dirac delta in the equatorial plane.

Numerical relativity predicts stationary toroidal sources\cite{Ansorg1} that can
be obtained by starting from a spheroidal topology for a sufficiently strong degree of
differential rotation; however, in the case of rigid rotation, these cannot be attained\cite{Ansorg2}. 

Added to the difficulty in finding suitable interiors, there are those arising
from the matching to the asymptotically flat exterior. For stellar models, it is an
overdetermined problem\cite{Mars-Seno}, so in general we cannot find an exterior that matches
a given interior. This seems to be the case for Wahlquist, where the derivations of
the impossibility of matching it with an asymptotically flat exterior  come
from analysis of the shape of its surface and involve approximations.

Within the field of approximations, one has to choose between accuracy and
closeness to the real physical problem that numerical methods provide,  and the density of information and greater flexibility for
theoretical work that analytic perturbation theory offers.

We would like to build an approximate solution of the Einstein equations which describes the gravitational field inside a ball of perfect fluid differentially rotating, and to match it, on the zero pressure surface, to an asymptotically flat approximate solution of the vacuum Einstein equations. 
The known exact solutions for this problem are either not physical \cite{Wahlquist, Senovilla} or they have some point of singularity \cite{Mars-Seno_2, Mars-Seno_3, Haggag}.

Although the metrics analysed fulfil the Einstein equations for differentially rotating
perfect fluids, none of them fulfils the set of requirements to become a physically
relevant solution.  The search for exact solutions for stationary axisymmetric gravitational fields coupled with differentially rotating perfect fluids remains open. 

In some previous papers \cite{CR,CMMR,CuGMR,MMR}, we studied this problem for rigid rotation; and now, we will use the same approximation scheme to study a differentially rotating perfect fluid. 

The scheme  we proposed consists of a slow rotation approximation on a
post-Minkowskian algorithm. We introduce two dimensionless parameters. One,
$\lambda$, measures the strength of the gravitational field, the other,
$\Omega$ which was a constant, measures the deformation of the matching surface due to fluid
rotation. 

If there is no rotation ($\Omega=0$), we are faced
with the post-Minkowskian perturbation to the Newtonian gravitational field
of a spherically symmetric mass distribution. Meanwhile,
Newtonian deformation of the source due to rotation is included in first-order $\lambda$ terms, up to some order in the rotation parameter. In this paper, with  differential rotation, the parameter 
used  is the rotation around the symmetry axis: $\Omega_0$. 

In the Newtonian formulation, a barotropic equation of state for the fluid implies that differential rotation can only depend on the cylindrical coordinate $\rho$ as a consequence of the Poincar\'e-Wavre theorem. 

In the barotropic case, the integrability conditions of the relativistic  Euler equation  imply that $\Phi= \Psi u_\varphi$ must be a function of $\Omega$, $\phi(\Omega)$. 
The choice of this function will determine the rotational model. 

A linear law has been used for some numerical results \cite{komatsu1989a,komatsu1989b}.  We will make this easy choice for our analytical model.  

In this paper, we keep terms of order less than or equal to $\Omega_0^5$ and  $\lambda^{3/2}$; that is, we have gone beyond a simple linear analysis but not so far as to compute strong non-linear effects.

\section{The metric and energy--momentum tensor}

The solution we are looking for is a stationary, 
axisymmetric and asymptotically flat space-time that admits a global system of spherical-like coordinates $\{t,r,\theta,\varphi\}$.

Our coordinates are adapted to the space-time symmetry, $\bm\xi=\partial_t$ and
$\bm\zeta=\partial_\varphi$, which are respectively the time-like 
and space-like
Killing vectors; so that the metric components do not
depend on coordinates
$t$ and $\varphi$, and the coordinates $\{r,\theta\}$ parametrize two-dimensional surfaces
orthogonal to the orbits of the symmetry group. Then we have:
\begin{eqnarray}
&\bm{g} = \gamma_{tt}\,\bm{\omega}^t{\otimes\,}\bm{\omega}^t
+\gamma_{t\varphi}(\bm{\omega}^t{\otimes\,}\bm{\omega}^\varphi+\bm{\omega}^\varphi{\otimes\,}\bm{\omega}^t)+
\gamma_{\varphi\varphi}\,\bm{\omega}^\varphi{\otimes\,}\bm{\omega}^\varphi
\nonumber\\
&\quad +\,\,\gamma_{rr}\,\bm{\omega}^r{\otimes\,}\bm{\omega}^r+
\gamma_{r\theta}(\bm{\omega}^r{\otimes\,}\bm{\omega}^\theta+\bm{\omega}^\theta{\otimes\,}\bm{\omega}^r) 
+\gamma_{\theta\theta}\,\bm{\omega}^\theta{\otimes\,}\bm{\omega}^\theta\,,
\label{eqmetrica}
\end{eqnarray}
where
$\bm{\omega}^t=dt$, $\bm{\omega}^r=dr$, $\bm{\omega}^\theta=r\,d\theta$, $\bm{\omega}^\varphi=r\sin\theta\,d\varphi$ 
is the Euclidean orthonormal co-basis associated with these coordinates.

Furthermore, coordinates $\{t,\,x=r\sin\theta\cos\varphi,\,y=r\sin\theta\sin\varphi,\,
z=\cos\theta\}$ associated with the spherical-like coordinates are harmonic
and the metric in these
coordinates tends to the Min\-kows\-ki metric in standard Cartesian 
coordinates for large values of the coordinate $r$.

We assume that the source of the gravitational field is a perfect fluid,
\begin{equation}
\bm{T} = \left(\mu + p\right)\bm{u}\otimes\bm{u}+p\,\bm{g}
\label{eqenermom}
\end{equation}
whose density  and pressure $p$ are functions of the $r$ and
$\theta$ coordinates. Moreover, we assume the fluid has no convective motion, so its velocity
$\bm{u}$ lies on the plane spanned by the two Killing vectors,
\begin{equation}
\bm{u} = \psi\left(\bm\xi + \omega\,\bm\zeta\right)\,,
\label{velocidad}
\end{equation}
where
\begin{equation}
\psi \equiv \left[-\left(\gamma_{tt}+2\omega\,\gamma_{t\varphi}\,r\sin\theta+\omega^2
\,\gamma_{\varphi\varphi}\,r^2\sin^2\theta\right)\right]^{-\frac12}
\label{eqnorma}
\end{equation}
is a normalization factor, i.e., $u^\alpha u_\alpha=-1$.

Let us consider the Euler equations for the fluid (or the energy--momentum tensor conservation law, which is equivalent):

\begin{equation}
\partial_a p = (\mu + p)\left( \partial_a\ln\psi-\Phi \partial_a \omega\right)
\qquad (a,b,\dots = r\,,\theta)\,.\label{eq5}
\end{equation}
where
\begin{equation}
\Phi\equiv \psi u_\varphi=-\frac{\gamma_{t\varphi}\,r\sin\theta+\omega \gamma_{\varphi\varphi}\,r^2\sin^2\theta}{\gamma_{tt}+2\omega\gamma_{t\varphi}\,r\sin\theta+\omega^2\gamma_{\varphi\varphi}\,r^2\sin^2\theta }
\label{defPhi}
\end{equation}
and $\omega$ is a function of $r$ and $\theta$.

If we consider a barotropic fluid, $\mu(p)$ then the integrability conditions for  (\ref{eq5}) are satisfied if and only if $\Phi$ is  a function of $\omega$ only, we call this function $\phi(\omega)$
to distinguish it from the  function defined in (\ref{defPhi}) which depends on the metric functions.

Therefore the solution of equations (\ref{eq5}) is implicitly defined by the equation
\begin{equation}
 \int^p \frac{dp'}{\mu(p')+p'}=\ln\psi-\chi\equiv \ln \zeta \label{Euler1}
\end{equation}
where 
\begin{equation}
\chi(\omega)\equiv \int^\omega  \phi(\omega') \, d\omega'\quad \mbox{and}\quad
 \zeta\equiv \psi e^{-\chi}\label{eqzeta}
\end{equation}
Since $p$ must be a function of $\zeta$, it will play the same role as $\psi$ played in the rigid rotation problem, i.e., it will determine the surfaces $p=\mbox{constant}$. For instance, the $p=0$ surface can implicitly be defined as:
\begin{equation}
p=0\quad \Longleftrightarrow \quad\zeta=\zeta_\Sigma\,,
\label{eqsuperficie}
\end{equation}
where $\zeta_\Sigma$ is an arbitrary constant.

Equation (\ref{Euler1}) and (\ref{eqsuperficie}) play an important role in our scheme. We use them to derive approximate expressions for the
pressure and the matching surface in a coherent way with the expansion for the metric we propose below.

Given an equation of state (EoS), we can integrate the left-hand side of (\ref{Euler1}) and even obtain explicit expressions for the pressure and density. For instance, a  linear equation of state,
$\mu+(1-n)p=\mu_0$, gives: 
\begin{equation}
 p=\frac{\mu_0}{n}\left(\left(\frac{\zeta}{\zeta_\Sigma}\right)^n-1\right)\quad\mbox{and}\quad \mu=\frac{\mu_0}{n}\left((n-1)\left(\frac{\zeta}{\zeta_\Sigma}\right)^n+1\right)\label{eqmuandp}
\end{equation}
whereas a  polytropic EoS,  
$p=a\mu^{1+1/n}$,
leads to:
\begin{equation}
 p=\frac{1}{a^n}\left(\left(\frac{\zeta}{\zeta_\Sigma}\right)^{\frac{1}{n+1}}-1\right)^{n+1}\quad\mbox{and}\quad \mu=\frac{1}{a^n}\left(\left(\frac{\zeta}{\zeta_\Sigma}\right)^{\frac{1}{n+1}}-1\right)^{n}
\end{equation}

As we said above, the function $\Phi$ can be written from (\ref{defPhi}) in terms of the metric components or, taking into account the integrability condition, as a function of only  $\omega$ and $\phi(\omega)$; by identifying both expressions we arrive at an equation which relates $\omega$ to the metric functions. Solving that equation for $\omega$ we will provide us with $\omega$, as a function of the metric and of the coordinates $r$ and $\theta$. 

To continue, we need to make some assumptions concerning the function $\phi(\omega)$. The simplest hypothesis is to choose a linear function:
\begin{equation}
 \phi=\frac{\omega_0-\omega}{\alpha}\label{eq14}
\end{equation}
This choice was already made \cite{komatsu1989a},\cite{komatsu1989b} within a numerical approach to differentially rotating polytropes\cite{Stergioulas}.

Now we can build up the function $\chi$: 
$$\chi=\int \phi d\omega=-\frac{(\omega-\omega_0)^2}{2\alpha }$$
where we have chosen the constant of integration in such a way that $\zeta =\psi$ at $\omega=\omega_0$, thereby recovering  previous results in rigid rotation.

\section{Approximation scheme}
 As in our previous work \cite{CMMR,CuGMR,MMR,CuMMR} on the rigid rotation problem, here we introduce a post-Minkowskian parameter, $\lambda$, and a dimensionless rotation parameter, $\Omega=\lambda^{-1/2}\omega r_0$,  where $r_0$ is the radius 
of the source in the  non-rotation limit. Then we can rewrite (\ref{eq14}) as:
 \begin{equation}
 \Phi=\lambda^{1/2} \frac{\Omega_0-\Omega}{r_0\alpha}.\label{HipPhi}
 \end{equation}
Moreover, we assume the following expansion of the metric components, (see \cite{CMMR}) (we will not use labels to distinguish between exterior or interior metrics  whenever it can be clearly understood to which of them we are referring). 
\begin{eqnarray}
&&\gamma_{tt} \approx -1 + \lambda f_{tt}\,,\quad
\gamma_{t\varphi} \approx\lambda^{3/2}\Omega_0f_{t\varphi}\,,\quad
\gamma_{\varphi\varphi} \approx 1+\lambda f_{\varphi\varphi}\,,
\nonumber\\
&&\gamma_{rr} \approx 1 + \lambda f_{rr}\,,\quad 
\gamma_{r\theta} \approx\lambda f_{r\theta}\,,\quad
\gamma_{\theta\theta} \approx 1+\lambda f_{\theta\theta}.\label{lambdaDepen}
\end{eqnarray}
We can determine $\Omega$ by solving the equation $\phi(\omega)=\Phi$, the expression (\ref{defPhi}) up to order $\lambda$ reads:
\begin{equation}
 \Phi\approx \lambda^{1/2} \Omega \frac{r^2\sin(\theta)^2}{r_0}+O(\lambda^{3/2})\label{Phirho}
\end{equation}
which means that, to the lowest order in $\lambda$, $\Omega$ will only depend on the cylindrical coordinate $\rho=r\sin(\theta)$, then from (\ref{HipPhi}) and (\ref{Phirho}) we obtain:
$$
\Omega\approx \frac{\Omega_0}{1+\alpha\rho^2}+\lambda F 
$$
where  function $F$, written in terms of the interior metric coefficients (\ref{lambdaDepen}), is:  
\begin{equation}F\approx -\alpha\Omega_0 r_0 \rho\frac{1}{1+\alpha\rho^2}\left(f_{t\varphi}+\frac{\rho}{r_0}\frac{f_{tt}+f_{\varphi\varphi}}{1+\alpha\rho^2}+\frac{\rho^3}{r_0^3}\frac{\Omega_0^2}{(1+\alpha\rho^2)^3}\right)\label{equF}
\end{equation}
Since we would like to follow the approximation scheme used in \cite{CMMR}, we need to assume that the constant $\alpha$ must be proportional to  $\Omega_0^2$ 
because it multiplies a second-order Legendre polynomial. Furthermore, as we want all our constants to be dimensionless, we redefine $\alpha$ as:
 $$\alpha\rightarrow \frac{\Omega_0^2}{r_0^2} \alpha$$
the expression for the angular velocity to the lowest order in $\lambda$ and up to $\Omega_0^5$ is:
\begin{equation}
 \Omega\approx
 \Omega_0\left(1-\alpha \Omega_0^2 \sin^2 \theta\eta^2+\alpha^2 \Omega_0^4 \sin^4\theta\eta^4\right),
\label{Omegapprox}
\end{equation}
substituting  this approximate expression for $\Omega$,
we obtain the following approximate expressions for $\psi$ and $\chi$, and omitting terms higher than  $\Omega_0^5$, we obtain:
\begin{equation}
 \chi\approx-\frac12 \lambda\alpha\Omega_0^4\frac{\rho^4}{r_0^4}\quad \mbox{and} \quad e^{-\chi}\approx 1+\frac12 \lambda\alpha\Omega_0^4\frac{\rho^4}{r_0^4}
\end{equation}
and:
\begin{equation}
 \psi\approx 1+\lambda\frac12 \left(f_{tt}+\Omega_0^2\frac{\rho^2}{r_0^2}-2\alpha\Omega_0^4\frac{\rho^4}{r_0^4}\right)\label{eqpsi}
\end{equation}
and finally, $\zeta$:
\begin{eqnarray}
 \zeta&\approx& 1+\lambda\frac12\left(f_{tt}+\Omega_0^2 \frac{\rho^2}{r_0^2}-\alpha \Omega_0^4 \frac{\rho^4}{r_0^4}\right)\equiv 1+\lambda\left(\frac12 f_{tt}+ \right.\nonumber\\[1em]
&&\left.\frac13 \Omega_0^2 \eta^2 (1-P_2)-\frac{4\alpha}{105} \Omega_0^4 \eta^4 (7-10 P_2+3P_4)\right)\label{eqzeta1}
\end{eqnarray}
where $\eta\equiv r/r_0$, and $P_l$ are the Legendre polynomials of $\cos(\theta)$. This approximate function is used to determine approximate expressions for the pressure, density and matching surface up to the order $\lambda$ and $\Omega_0^4$,
and consequently the energy--momentum tensor up to order $\lambda^{3/2}$ in our expansion parameter and  $\Omega_0^5$.

\section{First-order metric in harmonic coordinates}
As in our previous papers \cite{CMMR,CuGMR,MMR,CuMMR}, here we use the post-Minkowskian approximation scheme.
\begin{equation}
g_{\alpha \beta}=\eta_{\alpha \beta}+\lambda h_{\alpha \beta}
\end{equation}
In those references, the resulting equations and notation are explained .
\subsection{Linear exterior solution}
The inhomogeneous part of the linear exterior equations is zero, i.e.:
$$t_{\alpha\beta}=0,\quad  N_{\alpha\beta}=H_\alpha=0$$
and the equations to solve are:
\begin{eqnarray}
&&\triangle h_{\alpha\beta} = 0\,,\nonumber\\
[.6ex]
&&\partial^k (h_{k\mu} -\frac12
h\,\eta_{k\mu}\,) = 0\,.
\label{eqhomog}
\end{eqnarray}
We are going to assume equatorial symmetry and the same dependence of the metric on the expansion parameters  as in our previous work \cite{CMMR,CuGMR} ($M_n\propto \Omega_0^n$ , $J_n\propto \Omega_0^n$). So, the exterior metric up to order $\lambda$ and $\Omega_0^4$ 
can be written in terms of the spherical harmonic tensors as:
\begin{eqnarray}
&&\bm{h}\approx 2\lambda\sum_{l=0,2,4}\Omega_0^l\frac{M_l}{\eta^{l+1}}\left(\bm{T}_l+\bm{D}_l\right)
+2\lambda^{3/2}\sum_{l=1,3,5}\Omega_0^l\frac{J_l}{\eta^{l+1}}\,\bm{Z}_l
\nonumber\\
&&\quad\,\,\,\,+\,\lambda\sum_{l=0,2,4}\Omega_0^l\frac{A_l}{\eta^{l+3}}\,\bm{E}_{l+2}
+\sum_{l=2,4}\Omega_0^l\frac{B_l}{\eta^{l+1}}\,\bm{F}_{l}\,,
\label{eqsolinexthar}
\end{eqnarray}
 where
\begin{eqnarray}
&&\bm{T}_n\, \equiv P_n(\cos\theta)\,\bm{\omega}^t\otimes\,\bm{\omega}^t \quad (n\geq 0)\,,
\nonumber\\
&&\bm{D}_n \equiv P_n(\cos\theta)\,\delta_{ij}dx^i{\otimes\,}dx^j \quad (n\geq 0)\,,
\nonumber\\
&&\bm{Z}_n \,\equiv
P_n^1(\cos\theta)\,(\bm{\omega}^t\otimes\bm{\omega}^\varphi+\bm{\omega}^\varphi\otimes\bm{\omega}^t)
\quad (n\geq 1)\,,
\label{base1}
\end{eqnarray}
are spherical harmonic tensors and:\footnote{The definitions and notation used in this paper are the same as in \cite{CuGMR}, but not those in the former  work \cite{CMMR}}
\begin{eqnarray}
&&\bm{E}_n \equiv \frac12 n(n-1)\,\bm{H}_n+(n-1)\,\bm{H}^1_n -\frac12\,\bm{H}^2_n \quad (n\geq 2)\,,\nonumber\\
&&\hspace*{-1em}\bm{F}_n \equiv \frac13 n(2n-1)\,\bm{D}_n- \frac16 n(n+1)\bm{H}_n -\frac12(\bm{H}^1_n+\bm{H}^2_n) \quad (n\geq 1)
\label{basext}
\end{eqnarray}
are two suitable combinations of $\bm{D}_n$, and these other three spherical harmonic tensors:
\begin{eqnarray}
&&\bm{H}_n \equiv P_n(\cos\theta)\,(\delta_{ij}- 3e_i  e_j)dx^i{\otimes\,}dx^j \quad (n\geq 0)\,,
\nonumber \\
&&\bm{H}^1_n \equiv P_n^1(\cos\theta)\,(k_ie_j + k_je_i)dx^i{\otimes\,} dx^j \quad (n\geq 1)\,,
\nonumber \\
&&\bm{H}^2_n \equiv P_n^2(\cos\theta)\,(k_ik_j -m_im_j)dx^i{\otimes\,} dx^j \quad (n\geq 2)
\label{base2}
\end{eqnarray}
$k_i$, $e_i$ and $m_i$ stand for Euclidean unit vectors of standard cylindrical coordinates, 
$d\rho=k_i\,dx^i$, $dz = e_i\,dx^i$, $\rho\,d\varphi = m_i\,dx^i=\bm{\omega}^\varphi$ (this is the set  of spherical harmonic tensors we use to write covariant tensors of rank-2 in
this paper); $M_n$ and $J_n$ are the multi-pole moments in Thorne \cite{Thorne1} or, except for a constant, those in Geroch-Hansen \cite{Geroch,Hansen}, 
$A_n$ and $B_n$ are other constants which, unlike $M_n$ and $J_n$, are not
intrinsic, (gauge constants)  but they  are needed to solve the Lichnerowicz matching problem.
$\bm{E}_2$ has spherical symmetry (therefore it
must be included in the spherical symmetric linear solution in addition to
the mass monopole term). 
\begin{equation}
E_2^{ij} \equiv H_2^{\,\, ij}+H_2^{1\,\,ij}-\frac12 H_2^{2\,\,ij}=\delta^{ij}-3n^in^j\,,
\end{equation}
where $n^i$ is the Euclidean unit radial vector of standard spherical
coordinates.

Let us remark that  in this approximation, order $M_0$, $J_1$ and $A_0$ can be polynomials of second order in $\Omega_0^2$;
$M_2$, $J_3$, $A_2$ and $B_2$ can be linear functions of $\Omega_0^2$; and 
$M_4$, $J_5$ and $B_4$ are pure numbers.

To obtain the exterior metric, we must add the Minkowski part to the exterior solution for $\bm{h}$ (\ref{eqsolinexthar}), i.e.:
\begin{equation}
\bm{g}_{\rm ext}\approx -\bm{T}_0+\bm{D}_0+\bm{h}.\label{extg1}
\end{equation}

\subsection{Linear interior solution}
We will now find the interior solution for a fluid linear EoS.
To this end, we need an approximate expression of the energy--momentum tensor of the fluid. First of all, let us notice that  the density
$\mu_0=3\lambda/(4\pi r_0^2)$ is a quantity of order 
$\lambda$. Thus, taking into account (\ref{eqzeta}) and (\ref{eqmuandp}), it is easy to check that the pressure is of order
$\lambda^2$; so, to this order of approximation in $\lambda$, the density constant EoS and the linear one give the same energy momentum tensor. Therefore, the energy--momentum tensor (\ref{eqenermom}) contributes to the right-hand side of the Einstein equations by means of:
\begin{equation}
8\pi\,\bm{t} \approx
3\frac{\lambda}{ r_0^2}\left(\bm{T}_0+\bm{D}_0\right)
+6\frac{\lambda^{3/2}\Omega}{r_0^2}\eta\,\bm{Z}_1\,,
\label{impulsenerg0}
\end{equation}
if the terms of order equal to or higher than $\lambda^2$ are disregarded. Now we need to substitute $\Omega$ for its approximate expression in term of $\Omega_0$ at the lowest order in $\lambda$ and up to order $\Omega_0^5$ given in (\ref{Omegapprox});
and, also as we did in (\ref{eqzeta1}), we substitute $\sin^2\theta $ and  $\sin^4\theta $ for their expressions in terms of the Legendre polynomials $P_l(\cos\theta)$;
and finally we split the term containing  the tensor $\bm{Z}_1$ into the corresponding spherical harmonic tensors $\bm{Z}_1,\, \bm{Z}_3,\,\bm{Z}_5$.

Let us consider the following system of linear differential equations that corresponds to the linear post--Minkowskian approximation:
\begin{eqnarray}
&&\triangle h_{\alpha\beta} = -16\pi t_{\alpha\beta}\,,
\nonumber\\
[.6ex]
&&\partial^k (h_{k\mu} -\frac12 h\,\eta_{k\mu}\,) = 0\,,
\label{eqint}
\end{eqnarray}
where $\bm{t}$ is given by (\ref{impulsenerg0}).
A particular solution for the inhomogeneous part which is regular at the origin of the coordinate system $r=0$ is:
 \begin{eqnarray}
\bm{h}_{\rm inh}&=&-\lambda\eta^2\,(\bm{T}_0+\bm{D}_0)-\lambda^{3/2}\frac25\left[3\Omega_0 \eta^3\,\bm{Z}_1+\alpha \Omega_0^3 \eta^5 \left(\frac29 \bm{Z}_3-\frac67 \bm{Z}_1\right)\right.+\nonumber\\[1ex] && \left. \frac43 \alpha^2 \Omega_0^5 \eta^7 \left(-\frac{2}{7} \bm{Z}_1+\frac{1}{11} \bm{Z}_3-\frac{2}{91} \bm{Z}_5\right)\right]
\end{eqnarray}
Now, for the homogeneous part up to order $\lambda$ and $\Omega_0^5$, the regular solution at the origin with equatorial symmetry is:
\begin{eqnarray}
\bm{h}_{\rm hom}&\approx&
\lambda\left( \sum_{l=0,2,4} m_l \Omega_0^l \eta^l(\bm{T}_l+\bm{D}_l) +  \sum_{l=0,2,4} a_l \Omega_0^l \eta^l \bm{E^*}_l\right) + 
\nonumber\\[2ex]
&&  \lambda\sum_{l=0,2} b_{l+2}\Omega_0^{l+2} \eta^l \bm{F^*}_l
+\lambda^{3/2}\sum_{n=1,3,5}j_l \Omega_0^l \eta^l \bm{Z}_l
\label{inthom}
\end{eqnarray}
We have also introduced two new sets of spherical harmonic tensors:
\begin{eqnarray}
&&\bm{E^*}_0\equiv \bm{D}_0 , \quad  \bm{F^*}_0\equiv \bm{H}_0\nonumber\\[1ex]
&& \bm{E^*}_l \equiv \frac{l+1}{6}\left((4l+6)\bm{D}_l-l\bm{H}_l\right)-\frac12(\bm{H}^1_l+ \bm{H}^2_l)\quad (l\geq 1)\,,
\nonumber\\[1ex]
&& \bm{F^*}_l \equiv \frac12(l+1)(l+2)\,\bm{H}_l-(l+2)\,\bm{H}^1_l -\frac12\,\bm{H}^2_l
\quad (l\geq 2)\,,
\end{eqnarray}
which seem to be well suited to the interior problem, indeed better than those we used to write the linear exterior metric.

As before, in  this approximation order, $m_4$, $j_5$, $a_4$ and $b_4$  are pure numbers;  $m_2$, $j_3$, $a_2$ and $b_2$ are linear functions of $\Omega_0^2$;  and
$m_0$, $j_1$ and $a_0$ are quadratic functions of $\Omega_0^2$.

Finally, adding this homogeneous part to the inhomogeneous part, and to the Minkowski part, we obtain an approximate expression for the  interior
metric up to the order $\lambda^{3/2}$ and $\Omega_0^5$:
\begin{equation}
\bm{g}_{\rm int}\approx -\bm{T}_0+\bm{D}_0+\bm{h}_{\rm hom}+\bm{h}_{\rm inh}. \label{intg1}
\end{equation}

\subsection{Matching  surface and  energy--momentum tensor}
If we assume that the metric components are continuous 
on the matching surface, then we can use their exterior expressions  given by (\ref{lambdaDepen}) to make (\ref{eqsuperficie}) into  a true equation for this surface. So
we can search for a parametric form of the matching surface up to zero order in $\lambda$ and up to order $\Omega_0^4$ by making the following assumption:
\begin{equation}
r\approx r_0\left(1+\sigma_2 \Omega_0^2 P_2(\cos\theta)+\sigma_4 \Omega_0^4 P_4(\cos\theta)\right)\,.
\label{desenvradisup}
\end{equation}
where $\sigma_2$ and $\sigma_4$ must be determined from the equation $\zeta=$\emph{constant}.  Note that as for the previous constants in this approximation level, $\sigma_4$ is a constant but $\sigma_2$ could be a linear function of $\Omega_0^2$.
The function $f_{tt}$ can be read from  (\ref{eqsolinexthar}), i.e.:
$$f_{tt}=2 \sum_{n=0,2,4} \Omega_0^n\frac{M_n}{\eta^{n+1}}P_n(\cos \theta)$$
A simple but lengthy calculation leads to:
\begin{equation}
\sigma_2=\frac{1}{M_0}\left(M_2-\frac13\right)+\frac{4\Omega_0^2}{63 M_0^2}(9M_2-9M_2^2+6\alpha M_0-2)
\label{sigma2}
\end{equation}
\begin{equation}
\sigma_4=\frac{6+35M_0M_4-36M_2^2-6M_2-4\alpha M_0}{35M_0^2}
\label{sigma4}
\end{equation}
We can also obtain a similar expression for  $\zeta_\Sigma$ (\ref{eqsuperficie}) in terms of the exterior constants:
\begin{equation}
\zeta_\Sigma\approx 1+\lambda\left(M_0 +\frac{1}{3}\Omega_0^2-\Omega_0^4\frac{M_2-1+6M_2^2+4\alpha M_0}{15M_0}\right)\,.
\label{psis1}
\end{equation}

\subsection{Global solution to first order in $\lambda$}
Let us recall the matching conditions we are using in this paper:  the metric components and their first derivatives have to be continuous through the  hyper-surface
of zero pressure (matching surface). Imposing these conditions on metrics (\ref{extg1}) and (\ref{intg1}) on the
surface given by (\ref{desenvradisup}) and  bearing in mind the order of
approximation we are concerned with, a straightforward calculation leads to the following values for
the multipole moments:
\begin{eqnarray}
&&M_0=1+\frac{5}{12}\Omega_0^4,\quad M_2=-\frac12 +\frac17 \Omega_0^2\left(4\alpha-\frac53\right),\quad M_4=\frac17\left(\frac{15}{4}-\frac{2\alpha}{5}\right),\nonumber\\[1ex]
&& J_1=\frac25+\Omega_0^2\left(\frac13-\frac{8\alpha}{35}\right)+\Omega_0^4\left(\frac57-\frac{16\alpha}{21}-\frac{16\alpha^2}{105}\right),\quad \nonumber\\[1ex]
&& J_3=-\frac17+\frac{4\alpha}{315}+\Omega_0^2\left(-\frac{55}{294}+\frac{202\alpha}{735}+\frac{16\alpha^2}{1155}\right),\nonumber\\[1ex]
&& J_5=\frac{5}{42}-\frac{32\alpha}{1155}-\frac{16\alpha^2}{15015}
\end{eqnarray}
all the gauge constants $A_n$ and $B_n$ are zero. For the interior metric constants we obtain:
\begin{eqnarray}
&&m_0=3+\frac{5}{12}\Omega_0^4,\quad m_2=-1 +\frac17 \Omega_0^2\left(8\alpha-\frac{15}{2}\right),\quad m_4=-\frac{4\alpha}{35}\nonumber\\[1ex]
&& j_1=2+\Omega_0^2\left(\frac23-\frac{4\alpha}{5}\right)+\Omega_0^4\left(\frac56-\frac{32\alpha}{21}-\frac{16\alpha^2}{35}\right),\quad j_5=-\frac{16\alpha}{1155}(4+\alpha),\nonumber\\[1ex]
&& j_3=-\frac27+\frac{4\alpha}{35}+\Omega_0^2\left(-\frac{55}{147}+\frac{404\alpha}{735}+\frac{8\alpha^2}{105}\right)
\end{eqnarray}
and also all the gauge constants $a_n$ and $b_n$ are zero up to this order. The above expressions include the rigid rotation case, which is arrived at by making $\alpha=0$.  Furthermore, the resulting expressions enlarge the results given in \cite{CMMR},\cite{CuGMR} since they include corrections in $\Omega_0^4$ at the first order in $\lambda$.
\section{Conclusions}
In this paper we have extended our analytical approach to stationary axisymmetric global solutions 
for a barotropic rigidly rotating perfect fluid to a differentially rotating perfect fluid.
The integrability condition of the Euler equation for barotropic fluids demands that the function $\Phi\equiv \Psi u_\varphi$ must be a function of the differential rotation $\omega$.

We would like to comment on the function $\phi(\omega)$. Following proposals of other authors, \cite{komatsu1989a},\cite{komatsu1989b}, we have chosen a linear function and it seems to be well suited to  our approximation scheme if a simple assumption concerning the dependence of the constant $\alpha$ on $\Omega_0$ is made. Nevertheless, our scheme can also be compatible with a more general choice of this function. 

Let us consider the following choice:
\begin{equation}
\phi(\omega)=\frac{\omega_0-\omega}{\alpha}\left(1+\beta (\omega_0-\omega)^2+\cdots\right)\label{omeganew}
\end{equation}
In our scheme, even terms in $(\omega_0-\omega)$ must be omitted since  under a change in the sign of the rotation $$\gamma_{t\varphi}\rightarrow  -\gamma_{t\varphi}, \quad u_\varphi\rightarrow -u_\varphi,\quad \psi\rightarrow \psi$$
and then $\Phi \rightarrow -\Phi$.

Finally, introducing $\Omega$ and $\Omega_0$ in (\ref{omeganew}):
\begin{equation}
\phi(\Omega)=\frac{\lambda^{1/2}}{r_0\alpha}(\Omega_0-\Omega)\left(1+\beta\frac{\lambda}{r_0^2} (\Omega_0-\Omega)^2+O(\lambda^2)\right)
\end{equation}
Therefore, corrections to the linear function can be ignored unless we want to evaluate $\Omega$ to the first order in $\lambda$; that is, they change the function $F$ in (\ref{equF}) but do not modify the metric up to order $\lambda$. However, this choice of  $\phi(\omega)$ is compatible with our approximation scheme. Nevertheless, it will lead to a slightly different second-order metric, also including a new constant, $\beta$.

The choice of this function  as  a lineal function permits us to integrate the Einstein equations to each order, obtaining a series in two parameters: the post-Minkowskian parameter, $\lambda$, and the rotation on the symmetry axis,
$\Omega_0$.

With this choice, a new parameter appears: $\alpha$. When this parameter is zero, we obtain our previous results for the rigid rotation problem.

In this paper, we have obtained the global metric keeping terms of order less than or equal to $\Omega_0^5$ and  $\lambda^{3/2}$; that is, we have gone beyond a simple linear analysis but no so far as to compute strong non-linear effects. However, since the algorithm is implemented by an algebraic computational programme, our
results  can easily be enhanced, if so desired, by going farther in the approximation scheme.

In the Newtonian formulation, a barotropic fluid implies that differential rotation can only depend on the cylindrical coordinate $\rho$, as a consequence of the Poincar\'e-Wavre theorem. One interesting result is that we obtain this conclusion in the post-Minkowskian perturbation at the first order, i.e.:
$$
\Omega\approx f_1(\rho)+\lambda f_2(\rho,z)
$$ 
\section*{Acknowledgements}
AM gratefully acknowledge the \textit{Universidad de Salamanca }for the warm hospitality which facilitated this collaboration. We also want to thank our friend J.M.M. Senovilla for comments that have allowed us to improve the paper. Financial support to the authors for this work was provided by the Spanish people via the government award FIS2015-65140-P (MIN\-E\-CO/FE\-DER).

\end{document}